\documentstyle[12pt,aaspp4]{article}

\lefthead{Hirano \& Taniguchi}
\righthead{Submillimeter CO in L1157}

\begin{document}

\title{Submillimeter CO emission from shock-heated gas
in the L1157 outflow}

\author{Naomi HIRANO}
\affil{Department of Astronomical Science,
Graduate University for Advanced Studies,
Mitaka, Tokyo, 181-8588, JAPAN}
\authoremail{hiranonm@cc.nao.ac.jp}

\author{Yoshiaki TANIGUCHI}
\affil{Astronomical Institute, Graduate School of Science,
Tohoku University, Aramaki, Aoba, Sendai, 980-8578, JAPAN}

\begin{abstract}
We present the CO $J$=6--5, 4--3, and 3--2 spectra
from the blueshifted gas of the outflow
driven by the low-mass class 0 protostar in the L1157 dark cloud.
Strong submillimeter CO emission lines with $T_{\rm mb} >$ 30 K
have been detected at 63$''$ ($\sim$0.13 pc) south from the
protostar.
It is remarkable that the blue wings in the submillimeter lines
are stronger by a factor of 3--4 than that of the CO $J$=1--0
emission line.
The CO line ratios
suggest that the blueshifted lobe of this outflow consists of
moderately dense gas of $n$(H$_2$) = (1--3)$\times$10$^{4}$
cm$^{-3}$ heated to $T_{\rm kin}$ = 50--170 K.
It is also suggested that the kinetic temperature of the outflowing gas
increases from $\sim$80 K near the protostar
to $\sim$170 K at the shocked region in the lobe center,
toward which the 
largest velocity dispersion of the CO emission is observed.
A remarkable correlation between the kinetic temperature and
velocity dispersion of the CO emission along the lobe
provides us with
direct evidence that the molecular gas
at the head of the jet-driven bow shock
is indeed heated kinematically.
The lower temperature of $\sim$80 K measured at the other
shocked region near the end of the lobe
is explained if this shock is in a later
evolutionary stage, in which the gas has been cooled mainly through
radiation of the CO rotational lines.

\end{abstract}


\section{INTRODUCTION}

There is now increasing evidence that the extremely young outflows 
driven by class 0 protostars interact with the surrounding ambient 
gas and produce strong shock waves (e.g., 
\markcite{Bac92,Bac96}Bachiller
\& G\'{o}mez-Gonz\'{a}lez 1992; Bachiller 1996). 
Since the shocked gas is significantly
 heated and compressed, various chemical reactions that 
cannot proceed in the cold and quiescent dark clouds are triggered off
(e.g., \markcite{Bac97,Gar98}Bachiller \& P\'{e}rez Guti\'{e}rrez 1997;
Garay et al. 1998).
One prototypical and well studied example of an outflow with strong 
shocks is the well-collimated bipolar outflow in the L1157 dark cloud 
(e.g., \markcite{Ume92}Umemoto et al. 1992),
which is located at 440 pc from the Sun 
(\markcite{Vio69}Viotti 1969). 
The blue lobe of this outflow is known to 
consist of two limb-brightened cavities (\markcite{Gue96}Gueth, 
Guilloteau, \& Bachiller 1996); 
toward the tips of these two cavities, 
strong SiO and NH$_3$ (3, 3) emission lines which are believed to be good
 tracers of shocked molecular gas are observed
(\markcite{Gue98, Zha95,Taf95}Gueth, Guilloteau, \& Bachiller 1998;
Zhang et al. 1995; Tafalla \& Bachiller 1995). 
Recently, \markcite{Une99}Umemoto et al. (1999) 
have detected the highly 
excited NH$_3$ emission lines of ($J, K$) = (5, 5) and (6, 6) 
in the blue lobe. 
They have shown that the blueshifted gas in the L1157 outflow is 
heated to 130--140 K, which is at least by a factor of $\sim$10 
over the temperature of the quiescent gas
even though the outflow is excited by a low-luminosity source of only 
$\sim$11 $L_{\odot}$. 
However, their NH$_3$ observations have been performed with rather 
low angular resolution, $\sim$72$''$, and
therefore their results do not tell us 
the spatial distribution of the highly excited gas and the variation of 
the gas kinetic temperature in the lobe.

In this Letter, we present the results of our new CO $J$=6--5, 4--3, 
and 3--2 observations of the high-velocity gas in the blue lobe of 
the L1157 outflow.
Since the upper level of the CO $J$=6--5 transition lies 117 K 
above the ground, this emission line is expected to be a good 
tracer of the warm molecular gas.
Furthermore, ratios of this line to the 
lower energy transitions can be used to probe the 
kinetic temperature and density of the high-velocity gas.
Our spatial resolution of $\sim$14$''$ is high enough to explore
the spatial variation of the kinetic temperature in the outflow lobe
whose extent is $\sim$2$'$.

\section{OBSERVATIONS}

The observations of the CO $J$ = 6--5, 4--3, and 3--2 transitions
were carried out with the James Clerk Maxwell Telescope
(JCMT)\altaffilmark{1} in 2000 June.
The half-power beam widths (HPBW) of the telescope were 8$''$
at the frequency of CO 6--5 (691.473 GHz), 12$''$ at that of
CO 4--3 (461.041 GHz), and 14$''$ at the CO 3--2 frequency
(345.376 GHz).
The main beam efficiencies at 691, 461, and 345 GHz were
0.3, 0.5, and 0.62, respectively.
The CO 6--5 and 4--3 observations were carried out using the 
dual-channel SIS receiver RxW in the D-band and C-band, respectively,
with an atmospheric opacity at 225 GHz of $\sim$0.06.
The single-sideband system temperature was 7000--10000 K in the
D-band and was 2500--3000 K in the C-band.
We used the SIS receiver RxB3 in the single-sideband mode for
the CO 3--2 observations.
The system noise temperature was 520--570 K.
The backend was a digital autocorrelation spectrometer (DAS) set up
with a bandwidth of 500 MHz for the CO 6--5 observations and 
250 MHz for the CO 4--3 and 3--2 observations.
The frequency resolution of the DAS with 500 MHz bandwidth mode
was 625 kHz, corresponding to a velocity resolution of 0.27 km s$^{-1}$
at 691 GHz.
The frequency resolution of the DAS with 250 MHz mode was 312.5 kHz, 
which corresponds to the velocity resolution of 0.27 km s$^{-1}$ 
at 345 GHz and of 0.20 km s$^{-1}$ at 461 GHz.

We obtained spectra toward three positions along the blueshifted 
lobe of the CO outflow. 
The position offsets from the driving source, 
IRAS 20386+6751 (hereafter referred to as 
L1157 IRS), $\alpha$(1950) = 20$^{\rm h}$38$^{\rm m}$39.$^{\rm s}$6, 
$\delta$(1950) = 67$^{\circ}$51$'$33$''$, 
are (0$''$, 0$''$), (20$''$, -60$''$), and (45$''$, -105$''$), 
which were referred to as positions A, B, and C, respectively, by 
\markcite{Mik92}Mikami et al. (1992).
Positions B and C correspond to the tips of two CO cavities
 (\markcite{Gue96}Gueth et al. 1996).
The observed positions are shown in Figure 1 superposed on a 
CO ($J$=1--0) outflow map observed with the 45 m telescope of the
Nobeyama Radio Observatory (NRO) by Umemoto et al. (private
 communication).
We made five-point maps with 5$''$ spacing centered at the three 
positions, A, B, and C, in the CO $J$=6--5.
In the CO $J$=3--2 transition, we obtained the spectra along the line
through the three positions.
The sampling intervals were 6.3$''$ between positions A and B and 
6.8$''$ between positions B and C.
We used the position switching mode to obtain the spectra,
and integrated 10 minutes per position in the CO $J$=6--5,
5 minutes in the $J$=4--3, and 3 minutes in the $J$=3--2.
In the CO $J$=6--5 and $J$=4--3, the signals collected in two
channels were co-added in order to improve the signal-to-noise ratio.
The telescope pointing accuracy was checked by observing W75N and
NGC7538 IRS1, and was better than 3$''$.
We have obtained the spectra of W75N and NGC7538 IRS1, and 
used them as intensity calibrators.
The intensities and line shapes of the calibrator spectra observed
by us were consistent with those of the previously observed ones
in the CO 3--2 and 6--5 lines.
However, the intensity and line profile of the CO 4--3  line did not
agree well with the previous ones.
We also compared the CO 4--3 spectrum observed at position B 
with that observed in 1995 which is available in the JCMT data archive. 
We found that our peak temperature ($\approx$33 K) is lower than
the previous measurements ($\approx$40 K).
Therefore, the CO 4--3 intensity values of our measurements 
may be underestimated by $\sim$20\%.
Typical rms noises per channel were $\sim$0.3 K in $T_{\rm mb}$ for
CO 3--2, 0.9--1.2 K for CO 4--3, and 2.0--3.0 K for CO 6--5 spectra.

\altaffiltext{1}{The James Clerk Maxwell Telescope is operated by the Joint Astronomy Centre, on behalf of the Particle Physics and Astronomy Research Council of the United Kingdom, the Netherlands Organization for Science Research and the National Research Council of Canada}

\section{RESULTS AND DISCUSSION}

\subsection{Strong CO Emission Lines in the Blue Lobe}

In Figure 1, we show the CO $J$=6--5, 4--3, and 3--2 spectra
observed at positions A, B, and C along with those of the CO and 
$^{13}$CO $J$=1--0.
The CO $J$=6--5 spectra presented in Figure 1 were obtained
by convolving the data of five points centered at each position
with the 14$''$ Gaussian beam.
The spectra of the CO and $^{13}$CO $J$=1--0 were observed
with the NRO 45m telescope (the beam size was 15$''$)
by Umemoto et al. (private communication).
It is shown that the CO $J$=6--5, 4--3, and 3--2 emission lines are 
extremely strong ($T_{\rm mb} >$ 30 K) at position B
which is 63$''$ ($\sim$0.13 pc) away from the
driving source.
The peak temperatures of the CO $J$=6--5, 4--3, and 3--2 emission
lines observed at position B are 32, 33, and 37 K, respectively,
which are much higher than $T_{\rm mb} {\sim}$ 20 K in the
CO $J$=1--0. 

Figure 1 shows that the upper three transitions of CO (i.e.,
$J$=6--5, 4--3, and 3--2) have similar line profiles,
although the absorption dips in the $J$=6--5 lines are less prominent
than those of the $J$=4--3 and 3--2 ones.
The absorption dips are likely to be caused by the foreground
material with lower excitation because their velocities coincide
with the cloud systemic velocity of  $\sim$2.7 km s$^{-1}$
determined from the intensity peak of the $^{13}$CO emission.
A comparison of the profiles of the CO $J$=6--5, 4--3, and
3--2 lines with the $^{13}$CO $J$=1--0 ones implies that the 
upper three transitions of CO lines are emitted by the outflowing gas
even in the velocity ranges close to the cloud systemic velocity.
The spectra of CO $J$=1--0 show different 
characteristics from those of the upper transitions;
no absorption dips, smaller velocity extent at positions B and C, 
and less prominent blueshifted wing at position A.
The shapes of the CO $J$=1--0 spectra are also affected by the 
self-absorption because their peak velocities are 
blueshifted by $\sim$0.8 km s$^{-1}$
with respect to the cloud systemic velocity.

\subsection{Physical condition of the outflowing gas}

In Figure 2, we show the line intensity ratios of CO 6--5/3--2 and 
3--2/1--0 measured at position B.
We used the CO $J$=6--5 data convolved with the 14$''$ beam
in order to match their angular resolution with the CO $J$=3--2
and 1--0 data.
Because of the uncertainty of the CO $J$=4--3 intensity values,
the ratios against the CO $J$=4--3 are not used for the 
following analyses.
It is shown that the 6--5/3--2 line ratio is almost constant
to $\sim$0.7 throughout the blueshifted wings except 
the innermost velocities
of a few km s$^{-1}$ from the cloud systemic velocity.
On the other hand, the 3--2/1--0 ratio increases from $\sim$3 at
$V_{\rm LSR}$ = 0 km s$^{-1}$ to $\sim$6 at $V_{\rm LSR}$ = 
$-$7 km s$^{-1}$.
In the following analyses, we use the intensity weighted value of 
$\sim$4 as a 3--2/1--0 ratio at position B.
At positions A and C, variations of the 6--5/3--2 and 3--2/1--0
ratios as a function of the velocity show the same trends as
those of position B.
At these two positions, typical values of the 6--5/3--2 and 
3--2/1--0 ratios are $\sim$0.5 and $\sim$3, respectively.

In order to derive the physical parameters of the 
outflowing gas, we carried out Large Velocity Gradient (LVG)
statistical equilibrium calculations (\markcite{Gol74,Sco74}Goldreich \& Kwan 1974; Scoville \& Solomon 1974).
Detailed descriptions of the calculations are presented in Appendix A
of Oka et al. (1998).
We adopted a CO to H$_2$ abundance ratio $X$ of 10$^{-4}$.
The velocity gradient ${dv/dr}$ is assumed to be 
180 km s$^{-1}$ pc$^{-1}$, which is derived from
 the velocity dispersion of the CO $J$=3--2 line,
15 km s$^{-1}$, and the lobe width of 0.085 pc determined
from the half-power contour of the
CO $J$=1--0 map presented in Figure 1.

Figure 3 displays the CO 6--5/3--2 and 3--2/1--0 intensity
ratios
as a function of density and kinetic temperature.
It is shown that the line ratios observed at positions A and C
correspond to $n(\rm {H}_2) = 2 {\times} 10^4$ cm$^{-3}$ and 
$T_{\rm kin}$ = 75 K, and those at position B to 
$n(\rm {H}_2) = 3 {\times} 10^4$ cm$^{-3}$ and $T_{\rm kin}$ 
= 120 K.
If we assume a larger (smaller) value as the velocity gradient,
the solutions of the LVG calculations give us
slightly higher (lower) densities and lower (higher)
kinetic temperatures.
When we adopted the lobe widths determined from the interferometric
CO map of \markcite{Gue96}Gueth et al. (1996; 0.064 pc at position B
and 0.032 pc at position C), the CO $J$=3--2 line widths at 
positions B and C (15 km s$^{-1}$ and 7.5 km s$^{-1}$, respectively)
give a velocity gradient of ${dv/dr}$ $\sim$ 230 km s$^{-1}$ pc$^{-1}$.
In this case, we obtained 
$n(\rm {H}_2) = 3 {\times} 10^4$ cm$^{-3}$ and $T_{\rm kin}$ = 65 K
for positions A and C, and 
$n(\rm {H}_2) = 3 {\times} 10^4$ cm$^{-3}$ and 
$T_{\rm kin}$ = 100 K
for position B.
The kinetic temperature at position B is consistent with the
rotational temperature derived from the NH$_3$ ($J$, $K$) =
(3, 3) to (6, 6) data by Umemoto et al. (1999).
On the other hand, the kinetic temperature at position C derived
from the CO data is lower than the NH$_3$ rotational
temperature at the same position.
Because of the large beam size of 72$''$, the NH$_3$ spectra 
observed at position C by Umemoto et al. may have been
contaminated by the higher temperature component located at the
position B.

The LVG calculations indicate that the density of the outflowing gas
traced by the CO emission is typically in the range of (1--3)
$\times$ 10$^{4}$ cm$^{-3}$ and does not vary among the three
observed positions although the density at position B is slightly
higher than those of the other two positions.
Because of this rather low density, the CO transitions are not 
thermalized.
Especially, the CO $J$=1--0 transition falls in the parameter range of
superthermally excited or inversely populated.
Using the model parameters derived from the LVG analysis
and the observed total integrated intensities of the $J$=6--5, 3--2,
and 1--0 lines, we estimated the total energy released by the
rotational transitions of CO to be 1.2$\times$10$^{31}$ erg s$^{-1}$ at 
position B and 1.4$\times$10$^{30}$ erg s$^{-1}$ at position C.
The LVG model predicts that the $J$=6--5 line is the most luminous
CO transition at position B and accounts for $\sim$25\% of the
total CO luminosity.
On the other hand, the $J$=1--0 transition contributes only 0.5 \%
of the CO luminosity at the same position.
The $J$=4--3 and 3--2 lines account for 13\% and 5\% of the
total CO luminosity at this position, respectively.
At position C with lower gas temperature, the $J$=6--5, 4--3,
3--2, and 1--0 transitions contribute 26 \%, 23 \%, 11 \%, and
1 \% of the total CO luminosity, respectively.
It should be noted that the values estimated here (i.e., the energy
released by the CO lines and the relative luminosity of each
transition) depend little on the assumed $dv/dr$ value.

\subsection{Temperature variation along the blue lobe}

In Figure 4a, we show the position-velocity diagram of the CO $J$=
3--2 emission along the line through A, B, and C.
It is shown that the velocity dispersion of the 
CO $J$=3--2 line increases with increasing distance from the 
protostar toward position B, 
which corresponds to the tip of the CO cavity 
\markcite{Gue96}(Gueth et al. 1996).
This velocity pattern
can be explained if the cavity was
created by the passage of bow shock
(e.g., \markcite{Ben96,Smi97}Bence, Richer, \& Padman 1996;
Smith, Suttner, \& Yorke 1997).
The velocity dispersion suddenly decreases downstream
from position B, and slightly increases again toward position C.
Such a double-humped velocity dispersion shown in Figure 4a
can be explained in terms of multiple eruption events,
each of which having produced a bow shock.

In Figure 4b,  we plot the variation of the CO 3--2/1--0 intensity ratio
averaged over the blueshifted wing
($V_{\rm LSR}$ = $-$15 -- 1 km s$^{-1}$)
along the same line as Figure 4a.
The CO $J$=1--0 spectra used to obtain the 3--2/1--0 ratio
were reproduced from the mapped data obtained every 10$''$
spacing by resampling them with the 14$''$ Gaussian beam.
If we assume that the density of the outflowing gas is almost
constant along the lobe, $n(\rm{H}_2) = 2 {\times} 10^4$ cm$^{-3}$,
which is suggested from the ratios among the $J$=6--5, 3--2,
and 1--0 lines at positions A, B, and C,
the variation of the 3--2/1--0 ratio corresponds to 
that of the kinetic temperature (see Figure 3). 
At position A and the southernmost position of Figure 4b, the values
of the 3--2/1--0 ratio contain large uncertainties because the 
CO $J$=1--0 blueshifted wings are too weak.
If we exclude these two points at both ends, we can find that the
variation of the 3--2/1--0 ratio, that is, the variation of the
kinetic temperature correlates well with that of the
velocity dispersion of the CO $J$=3--2 emission.
Figure 4b shows that the temperature of the outflowing gas,
which is $\sim$80 K near the central source, 
increases toward position B
and reaches up to $T_{\rm kin}$ = 170 K
(under the assumption of $dv/dr$ = 180 km s$^{-1}$)
at the position where the largest velocity dispersion is observed.
Such a remarkable agreement between the velocity dispersion
and kinetic temperature can be explained by the model of 
jet-driven bow shock (e.g., \markcite{Mas93}Masson \& Chernin 1993), 
in which both momentum and energy
input from the jet to the swept-up medium are maximal
behind the bow shock.
A model calculation including the balance of heating and cooling 
(\markcite{Hat99}Hatchell, Fuller, \& Ladd 1999) has shown that 
the kinetic energy loss at the bow shock is sufficient to heat the
swept-up molecular gas and produces the temperature gradient
that increases from the protostar towards the bow shock.
At the downstream side of position B, the temperature once
drops to $\sim$50 K and increases again to $\sim$80 K at 10$''$
upstream side of position C.
The temperature increase toward the shock near position C
is less significant as compared to that toward the shock at position B.

Toward the shock at position B, 
the emission from the ro-vibrational H$_2$
lines in the near-infrared and that from the pure rotational lines
of H$_2$ in the mid-infrared have been detected (\markcite{Dav95,
Cab98}Davis \& Eisl\"{o}ffel 1995; Cabrit et al. 1998).
Since the required temperatures to excite the ro-vibrational and 
pure rotational lines of H$_2$ are $\sim$2000 K and 700-800 K,
respectively, the detection of these lines implies that the molecular
gas in the shocked region is heated to above 2000 K.
Because of the very short cooling timescale of such hot molecular
gas, $\sim$10 yr (e.g., 
\markcite{Neu95}Neufeld, Lepp, \& Melnick 1995),
the presence of hot gas means that the energy input was made very
recently in the shocked gas at position B.
On the other hand, no H$_2$ emission appears to be associated with the
other shock near position C, suggesting that the major energy input
process has been already suspended there.
By using the total CO luminosity expected at position B, 1.2$\times$
10$^{31}$ erg s$^{-1}$, and the H$_2$ column density derived at the
same position, $\sim$6$\times$10$^{21}$ cm$^{-2}$, we can estimate
the cooling rate per H$_2$ molecule by the rotational CO line emission
to be $\sim$3$\times$10$^{-25}$ erg s$^{-1}$.
If we assume that the rotational lines of CO are the major coolant of
the gas with temperature less than several hundred K
as in the case of the gas with $T_{\rm kin} <$ 40 K
(\markcite{Gol78}Goldsmith \& Langer 1978), 
it takes $\sim$1000 yr to cool the gas from the temperature 
at position B, 120 K, to 75 K derived at position C.
Since this cooling time scale is comparable to the difference of 
time scale between the two shocks 
(\markcite{Gue96}Gueth et al. 1996), 
the lower gas kinetic temperature 
at the shock near position C can be explained if this shock is in a
later evolutionary stage, in which the gas has been cooled through
radiation of the CO rotational lines.
Theoretical studies of the thermal balance in dense 
[$n$(H$_2$)$\gtrsim$10$^3$ cm$^{-3}$]
molecular clouds predicted that H$_2$O and O {\sc i} would be
important gas coolants (e.g., \markcite{Neu95}Neufeld et al. 1995).
However, because of the large dipole moment of the water molecule,
the H$_2$O lines cannot be the major coolant of the gas unless the
density is higher than 10$^{6}$ cm$^{-3}$.
On the other hand, contribution of the O {\sc i} fine structure lines to the
cooling of the gas with  $T_{\rm kin} >$ 100 K could be comparable
to that of the CO rotational lines.
However, the
gas cooling rate through O {\sc i} emission is expected to 
decrease rapidly
with decreasing gas temperature, because the upper energy state of the
lowest transition, $^{3}P_{1}$, is 228 K above the ground.

The total energy released by the rotational CO lines estimated
toward position C is one order of magnitude lower than that of
position B.
Therefore, the cooling time scale at position C where the
gas kinetic temperature is $\sim$80 K is estimated to be 
of the order to 10$^4$ yr,
which is comparable to the dynamical time scale of the outflows
driven by the Class 0 sources.
This implies that the kinetic temperature of the outflowing gas 
can be maintained at several tens of K, which is a factor of 3--5
higher than the ambient temperature, during the life time of the
Class 0 stage

\acknowledgments

We would like to thank the staff of the JCMT for the operation of our 
observations, and Dr. R. Phillips for assisting us with our observations
and data reduction. 
We acknowledge T. Oka for providing the LVG calculation
code of good performance, and T. Umemoto for providing us with 
the CO and $^{13}$CO $J$=1--0 data observed with the NRO 45m
telescope.
We also acknowledge the helpful comments of an anonymous referee
whose suggestions improved the paper.

\clearpage

\figcaption[fig1.eps]{CO and $^{13}$CO spectra observed at
 positions A (0$''$, 0$''$), B (20$''$, -60$''$), and C (45$''$, -105$''$).
The offsets are in arcseconds from the position of IRAS 20386+6751 [$\alpha$(1950) = 20$^{\rm h}$ 38$^{\rm m}$39.$^{\rm s}$6, $\delta$(1950) = 67$^{\circ}$51$'$33$''$].
The vertical line indicates the cloud systemic velocity of
$V_{\rm LSR}$ = 2.7 km s$^{-1}$ determined from the peak of
the $^{13}$CO emission.
The map of CO $J$=1--0 outflow was obtained with the NRO 45 m
telescope by Umemoto (private communication). 
The filled squares corresponds to the positions A, B, and C.
Solid line denotes the cut along which the CO $J$=3--2 was observed.
Dashed contours of the CO $J$=1--0 map
indicate the distribution of the blueshifted emission 
(-5 to 1 km s$^{-1}$) and solid contours show the redshifted 
emission (4.5 to 10 km s$^{-1}$). 
Contours are every 2.88 K km s$^{-1}$ (1.5 $\sigma$) with the
lowest contour at 5.76 K km s$^{-1}$ (3 $\sigma$).\label{fig1}}

\figcaption[fig2.eps]{Intensity ratios between CO $J$=6--5 and
$J$=3--2 (open squares) and CO $J$=3--2 and $J$=1--0 (filled
circles) measured at position B plotted as a function
of the LSR velocity.
The broken line denotes the cloud systemic velocity of
$V_{\rm LSR}$ = 2.7 km s$^{-1}$.
 Vertical error bars give 1$\sigma$ errors. \label{fig2}}

\figcaption[fig3.eps]{Curves of constant line ratios of
CO 6--5/3--2 (broken lines) and 3--2/1--0 (solid lines) as a 
function of $T_{\rm kin}$ and $n$(H$_2$) for $dv/dr$ = 180 km s$^{-1}$.
Contour intervals of the CO 6--5/3--2 and 3--2/1--0 ratios are
0.1 and 0.5, respectively.
 \label{fig3}}

\figcaption[fig4.eps]{(a) Position-velocity diagram of 
the CO $J$=3--2
emission along the major axis of the blue lobe (solid line in Fig. 1).
The lowest contour level and the contour spacings are 2 K.
(b) A plot of the variation of the CO 3--2/1--0 intensity ratio of the
blueshifted component ($V_{\rm LSR}$ range is from $-$15 to 
1 km s$^{-1}$) along the same line as Fig. 4a.
The scale on the upper side of the plot gives the corresponding
kinetic temperature for $dv/dr$ = 180 km s$^{-1}$ and
$n$(H$_2$) = 2$\times$10$^4$ cm$^{-3}$.
Horizontal error bars give 1$\sigma$ errors.
 \label{fig4}}

\end{document}